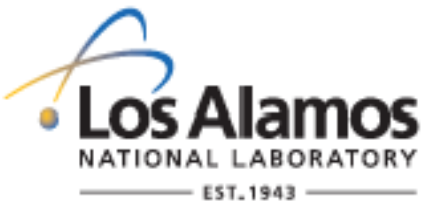

# MCNP6 Study of Spallation Products from 500 MeV p + $^{136}$Xe

Stepan G. Mashnik

*Los Alamos National Laboratory, XCP-3, MS A143, Los Alamos, NM 87545, mashnik@lanl.gov*

## INTRODUCTION

MCNP6 [1] is used in various applications involving reactions induced by neutrons and other low-energy projectiles, but also may be applied to particle- and heavy-ion collisions at relativistic energies. An improved version of the Cascade-Exciton Model (CEM) as implemented in the code CEM03.03 [2], is the main "workhorse" (event generator) used by MCNP6 as its default option to describe reactions induced by nucleons, pions, and photons at energies up to several GeV. It is critical that it be able to describe such reactions as well as possible; therefore, it is extensively validated and verified against available experimental data and calculations by other models (see, e.g., [3] and references therein). So far, for proton-induced reactions at intermediate energies, MCNP6 has been compared mostly with different particle spectra measured from various reactions as well as with yields of products from heavy actinide and from relatively light nuclei-targets [3] and much less with data on isotope-production yields from reactions on intermediate nuclei with mass numbers $A$ around and above ~100, which are usually much more difficult to calculate with any models (see [4] and references therein). To remedy this deficiency, we tested MCNP6 using CEM03.03 against the product yields from the 500 MeV p + $^{136}$Xe interactions measured recently at GSI in inverse kinematics [5]. The production of final residual nuclei in such reactions is of great interest for a number of applied and academic problems related to transmutation of nuclear wastes, radioactive beam facilities, propagation of cosmic radiation, understanding the reaction mechanisms leading to the production of highly-excited nuclei and to dissipation of kinetic energy in internal excitation energy of the nucleus and in the de-excitation process of such hot nuclei, to name just a few [5].

## RESULTS

The easiest way to calculate with MCNP6 production cross sections from a thin target is to use its GENXS option [6]. As this option was developed by Dick Prael specifically for MCNP6 and was not available in the older MCNP5 and MCNPX, it is not known very well yet to the MCNP community of users. This is why in Appendixes A and B we provide the complete MCNP6 input files for this problem (note, that the GENXS option requires a second, auxiliary input file, in addition to the main MCNP6 input file). Ref. [6] together with the MCNP6 User's Manual will help the readers to understand very easily all parameters in both input files.

The cross sections of all measured [5] isotopes are compared with our MCNP6 results in Fig. 1. We see that MCNP6 using CEM03.03 reproduces reasonably well most of the measured cross sections. For spallation products heavier than Cs but not too far from the target, the agreement between practically all calculated and measured cross sections is good, better than within a factor of two.

However, for the production of isotopes of Ba, with a charge number Z=56, greater than the charge number on the target equal to 54, by two units, we see an underestimation of almost an order of magnitude. This is a known problem of practically all models based on IntraNuclear Cascade (INC) to describe products with the charge number Z+2 in comparison with charge number Z of the target-nuclei from various intermediate-energy spallation reactions (see, e.g., Ref. [7]). Such a disagreement may be caused by several different details of INC-based models, but it is not completely understood at present, remaining still an open question of practically all INC-based Monte Carlo codes: We see here room for further improvement of nuclear reaction models and of our event generators used in different transport codes.

For the production of lighter isotopes, far from the target-nucleus, starting from Tc, we see another quite big underestimation of all measured cross sections by MCNP6: The further from the target, the bigger is the underestimation. The reason of this second disagreement is different from the problem with the Ba isotopes discussed above. It is related to the fact that the further the products are from the target, the bigger become the contributions to the production of such isotopes from nuclear reaction mechanisms not accounted by CEM03.03, like multifragmentation [8] or/and fission-like binary decays as described by the code GEMINI of Charity *et al.* [9].

The problems of nuclear reaction models describing well products far from the target-nuclei, in the deep spallation and fragmentation regions, are well known and have been studied with CEM for several other similar nuclear reactions (see [4, 10] and references therein). We cannot confirm that at present these problems are completely understood. We do not know unambiguously how big should be the contributions from multifragmentation and/or from fission-like binary decays to various nuclear reactions and how these contributions change with the incident energy and the mass and charge numbers of the target-nuclei, as well as of the products. For these reasons, the CEM03.03 event generator used at present by default in MCNP6 does not consider multifragmentation and fission-like binary decays. Further investigations and further improvement of our event generators are needed in order to predict well products from intermediate- and high-energy nuclear reactions far from the target-nuclei, in the deep spallation and fragmentation regions.

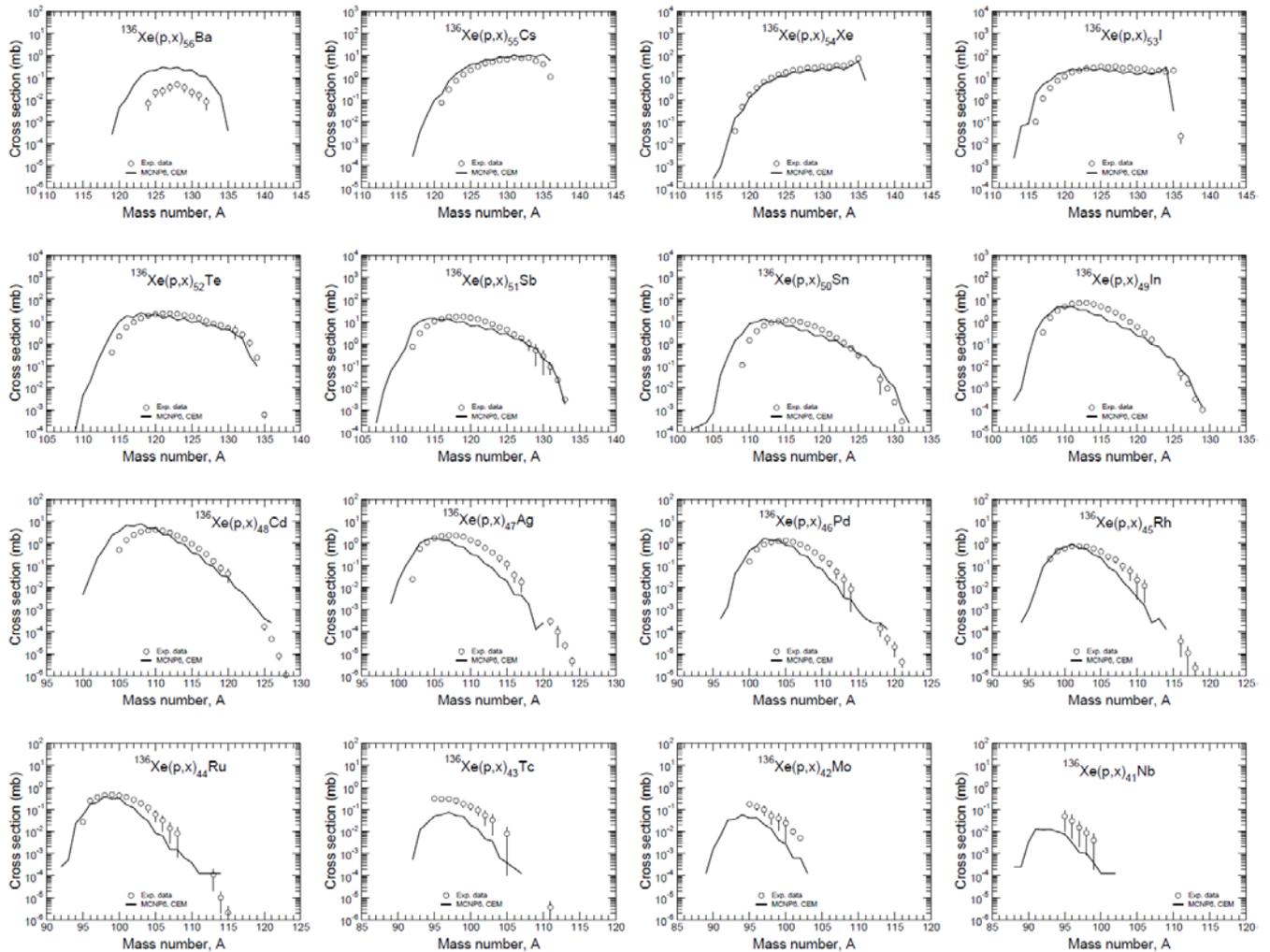

Fig. 1. Isotopic cross sections of all measured at GSI by Giot at al. [5] products from the reaction 500 MeV p + $^{136}$Xe (circles) compared with results by MCNP6 [1] using the CEM03.03 event generator [2].

## CONCLUSIONS

Isotope production cross sections from the 500 MeV p + $^{136}$Xe interactions measured recently at GSI in inverse kinematics have been simulated with MCNP6 using the CEM03.03 event generator. MCNP6 reproduces reasonably well most of the measured cross sections. However, for the production of isotopes with a charge number greater than the charge number on the target by two units, we see an underestimation of almost an order of magnitude. We found another quite big underestimation for the production of lighter isotopes, far from the target-nucleus, starting from Tc. Further investigations and further improvement of our event generators with a possible consideration of multifragmentation and fission-like binary decays are needed in order to predict well products from intermediate- and high-energy nuclear reactions far from the target-nuclei, in the deep spallation and fragmentation regions.

## ACKNOWLEDGMENTS

We acknowledge Dr. Lydie Giot for her interest in our CEM simulations at the initial stage of her acquirement of these data at GSI during 2006 and for some interesting discussions. We thank Dr. Roger Martz for a careful reading of our manuscript and several useful suggestions.

This work was carried out under the auspices of the National Nuclear Security Administration of the U.S. Department of Energy at Los Alamos National Laboratory under Contract No. DE-AC52-06NA25396.

## APPENDIX A: MCNP6 main input file

```
MCNP6 test: p + Xe136 by CEM03.03 at 500 MeV
 1  1  1.0   -1 2 -3
 2  0       -4 (1:-2:3)
 3  0        4

c ---------------------------------------------------------------
 1 cz  4.0
 2 pz -1.0
 3 pz  1.0
 4 so 50.0

c ---------------------------------------------------------------
 m1  54136 1.0
 sdef erg = 500 par = H dir = 1 pos = 0 0 0 vec 0 0 1
 imp:h 1 1 0
 phys:h 1000
 mode  h
c LCA  8j 1    $ use CEM03.03 (default)
 tropt genxs inxc96  nreact on  nescat off
c ---------------------------------------------------------------
 print 40 110 95
 nps 10000000
```

## APPENDIX B: MCNP6 auxiliary input file

```
MCNP6 test: p + Xe136 by CEM03.03 at 500 MeV
0 0 1 /
Cross Section Edit
0 0 9 /
1 5 6 7 8 21 22 23
```